\documentclass[onecolumn,showpacs,preprintnumbers,amsmath,amssymb]{revtex4}

\usepackage{float}
\usepackage{graphicx}
\usepackage{bm}
\usepackage{epstopdf}
\usepackage{color}

\begin{document}

\title{Generalized Neumann boundary condition for the scalar field}

\author{J.C. Fernandes}
\email{siamdhearks@gmail.com}
\affiliation{IFQ - Universidade Federal de Itajub\'a, Av. BPS 1303, Pinheirinho, Caixa Postal 50, 37500-903, Itajub\'a, MG, Brazil.}

\author{J.P. Ferreira}
\email{ferreira_jp@unifei.edu.br}
\affiliation{IFQ - Universidade Federal de Itajub\'a, Av. BPS 1303, Pinheirinho, Caixa Postal 50, 37500-903, Itajub\'a, MG, Brazil.}

\author{F.E. Barone}
\email{frederico.barone@gmail.com}
\affiliation{Niter\'oi, RJ, Brazil}

\author{F.A. Barone}
\email{fbarone@unifei.edu.br}
\affiliation{IFQ - Universidade Federal de Itajub\'a, Av. BPS 1303, Pinheirinho, Caixa Postal 50, 37500-903, Itajub\'a, MG, Brazil.}

\author{G. Flores-Hidalgo}
\email{gfloreshidalgo@unifei.edu.br}
\affiliation{IFQ - Universidade Federal de Itajub\'a, Av. BPS 1303, Pinheirinho, Caixa Postal 50, 37500-903, Itajub\'a, MG, Brazil.}

\author{L.H.C. Borges}
\email{luizhenrique.borges@ufla.br}
\affiliation{Departamento de F\'\i sica, Universidade Federal de Lavras, Caixa Postal 3037, 37200-900 Lavras-MG, Brazil.}

\begin{abstract}
In this paper, we explore the Klein–Gordon field theory in $(D+1)$ dimensions in the presence of a $(D-1)$-dimensional hyperplanar $\delta$-like potential that couples quadratically to the field derivatives. This model effectively generalizes the Neumann boundary condition for the scalar field on the plane, as it reduces to this condition in an appropriate limit of the coupling parameter. Specifically, we calculate the modifications to the Feynman propagator induced by the planar potential and analyze the interaction energy between a stationary point-like source and the potential, obtaining a general and exact expression.

We demonstrate that, under certain conditions relating the field mass and the coupling constant to the external potential, the vacuum state becomes unstable, giving rise to a pair-creation phenomenon that resembles the Schwinger effect in quantum electrodynamics.
\end{abstract}

\maketitle

\section{Introduction}
\label{Intro}

The interaction between point-like objects and material boundaries has been extensively studied in the context of Van der Waals interactions \cite{Feinberg,Lifshitz,Dzyaloshinskii,Sandoghdar} and the zero-point energy formalism \cite{MANSOM,Wylie,Wylie2,Sukenik} over the years. In particular, Casimir and Polder investigated the interaction between a neutral but polarizable atom and a perfectly conducting plane, incorporating retardation effects on London–Van der Waals forces within the framework of perturbative quantum electrodynamics \cite{Casimir1,KMilton,FarinaSomeAspects}. Their work led to an elegantly simple result that depends only on fundamental constants of nature, namely the speed of light ($c$) and the reduced Planck constant ($\hbar$), the atom's polarizability, and the distance between the atom and the plane. By utilizing the concept of zero-point energy, Casimir and Polder re-derived this result by calculating the variations in electromagnetic zero-point energy induced by the presence of the atom and the conducting surface \cite{Casimir2}.

The field-theoretical description of material boundaries has played a significant role in the literature, particularly through the application of quantum field theory to condensed matter physics. Notable examples include the study of topological insulators and artificial gauge field materials, as discussed in references \cite{DIRACMATERIALS, GAUGEMATERIALS}. This subject has garnered substantial interest for both experimental and theoretical reasons, as typical experimental setups often involve material boundaries that must be accurately incorporated into theoretical models. In this context, several key investigations stand out. These include studies of the interactions between semi-transparent mirrors and point-like charges \cite{  LHCBAFFFAB, HLOLHCBFAB, LHCAFF, LeeWickEspelhoSemiTransparente}, the interaction between an atom and a $\delta$-like mirror \cite{PKKM, Russo}, and the Casimir energy arising from the presence of two semi-transparent mirrors \cite{BorUM, KimballA, BHR1992, BordKD, NRVMH, NRVMMH2, PsRj, FoscoLousada, CapaFoscoLousada, Caval, FABFEB2, Mostepanenko,MostepanenkoLivro,Milton, RBEF2009}, among others. These studies underscore the importance of understanding material boundaries in both theoretical frameworks and experimental applications.

In this context, theories with quadratic couplings involving Dirac delta potentials concentrated on surfaces provide a powerful framework for investigating the role of material surfaces and/or interfaces in the presence of quantum fields. These $\delta$-like external potentials have proven to be effective tools for generalizing boundary conditions, especially when an appropriate limit is taken for the coupling constant. This approach has been successfully applied in the scalar field context to recover the Dirichlet boundary condition \cite{BHR1992,GTFABFEB} and the scalar MIT boundary condition \cite{walisson}. For the electromagnetic case, $\delta$-like planar potentials can similarly describe perfect conducting plates in the strong coupling limit, as shown in \cite{FABFEB}. This same approach allows us to explore aspects of the interaction between a stationary point-like charge and the plane, investigating the scalar version of Casimir–Polder interactions. A natural question that arises in this scenario concerns how a $\delta$-like external potential can be used to generalize Neumann boundary conditions on a surface \cite{argentinos}, and how such a potential would interact with a point-like source.


This paper is devoted to investigate this question. Here, we explored the scalar field theory in $(D+1)$ spacetime dimensions with the metric $\eta^{\mu\nu} = \text{diag}(+1, -1, -1, \ldots, -1)$ which couples quadratically to a planar $\delta$-like external potential involving field derivatives. In Section \ref{Prop}, we compute the exact correction to the Feynman propagator due to the presence of the planar potential and observe that, when the appropriate limit of the coupling constant is taken, the propagator of the theory—as well as any classical field solution—satisfies Neumann boundary conditions. This result aligns with the approach and findings of \cite{GTFABFEB} for Dirichlet conditions. In Section \ref{interaction}, we analyze the interaction energy between the planar $\delta$-like potential and a point-like source. Interestingly, we find that the image charge has the same sign as the original charge, contrasting with the Dirichlet-like scenario discussed in \cite{GTFABFEB}. Moreover, we demonstrate that, under the influence of the stationary point-like source and specific conditions relating the scalar field mass and the coupling parameter, the vacuum state of the theory becomes unstable. This instability leads to particle pair production, indicating that the vacuum-to-vacuum amplitude vanishes in the long-time limit during which the external current is present. Finally, Section \ref{conclusions} presents our conclusions and final remarks.

In this work, four-vectors are denoted by \( x = (x^0, \mathbf{x}) \),  and we consider a delta function localized on the plane $x^{D} = a$. The spatial coordinates are written as \( \mathbf{x} = (\mathbf{x_\parallel}, x^D) \), where \( \mathbf{x_\parallel} = (x^1, \dots, x^{D-1}) \) represents the coordinates parallel to the plane $x^{D}=a$, while \( x^D \) denotes the coordinate perpendicular to it.

\section{The lagrangian and its associated propagator}
\label{Prop}

In this section, we consider a real Klein-Gordon field in the presence of an external planar \((D-1)\)-dimensional \(\delta\)-like potential located on the plane \( x^D = a \). We derive an exact expression for the Feynman propagator associated with the underlying model. The potential couples quadratically to the field derivative in the form \( M^{-1}(n^\mu \partial_\mu \phi)^2 \delta(x^D - a) \), where \( M \) is the coupling parameter with the dimension of mass, and \( n^\alpha = (0, 0, \dots, 1) \) is a spacelike four-vector perpendicular to the plane \( x^D = a \).   
The model is governed by the following Lagrangian density 
\begin{equation}
\label{model}
\mathcal{L} = \frac{1}{2}\partial_\mu \phi \partial^\mu \phi - \frac{1}{2}m^2\phi^2 - \frac{\left(n^\mu\partial_\mu \phi \right)^2}{2M}\delta(x^D-a)+J\phi,
\end{equation}
where $m$ stands for the field mass with an implicit small imaginary part $m\to m-i\epsilon$ ($\epsilon>0$) in order to warrant convergence of functional integrals \cite{ashokPath}, and is an external source linearly coupled to the field.

It is worth mentioning that the coupling in the model (\ref{model}) is not equivalent to that presented in \cite{argentinos}, which involves the derivative of the Dirac delta function.

The propagator $G\left(x,y\right)$ of theory (\ref{model}) satisfies the following differential equation
\begin{equation}
\label{opera1}
\left\{\partial_\mu\partial^\mu + m^{2}-\frac{1}{M}n^\mu\partial_{\mu}\left[\delta(x^D-a)n^\nu\partial_{\nu}\right]\right\}G\left(x,y\right)=\delta^{D+1}\left(x-y\right),
\end{equation}
where $n^\mu\partial_{\mu}\left[\delta(x^D-a)n^\nu\partial_{\nu}\right]G(x,y)=n^\mu\partial_{\mu}\left[\delta(x^D-a)n^\nu\partial_{\nu}\;G(x,y)\right] $.

We can write the solution of Eq. (\ref{opera1}) in terms of a Beth-Salpeter-like equation \cite{GTFABFEB,LHCBAFFFAB,Caval} 
\begin{equation}
\label{prop1}
G(x,y)=G_0(x,y)+\frac{1}{M}\int d^{D+1}x'\;G(x,x')n^\mu \partial'_\mu\Big[\delta(x'^D-a)n^\nu\partial'_{\nu}\Big]G_0(x',y),
\end{equation}
with $G_{0}\left(x,y\right)$ standing for the Feynmann propagator, which satisfies
\begin{equation}
    \label{feynmanproptotal}
    \Big(\partial_\mu\partial^\mu+m^2\Big)G_{0}(x,y)=\delta^{D+1}(x-y).
\end{equation}

Due to the translational symmetry of the operator in the left side of Eq. (\ref{opera1}) in the $(D-1)$ coordinates parallel to the plane $x^D=a$, we can expand the propagators $G(x,y)$ and $G_0(x,y)$ in a Fourier integral over these parallel coordinates, as follows
\begin{eqnarray}
\label{rprop1}
G(x,y) &=& \int \frac{d^{D}p_{\parallel}}{(2\pi)^D}\;\mathcal{G}(p_\parallel;x^D,y^D) e^{-i p_{\parallel}.(x_{\parallel} - y_{\parallel})}  ,\\
\label{rprop2}
G_0(x,y) &=& \int \frac{d^{D}p_{\parallel}}{(2\pi)^D}\;\mathcal{G}_0(p_\parallel;x^D,y^D)e^{-i p_{\parallel}.(x_{\parallel} - y_{\parallel})}.
\end{eqnarray}
We shall call $\mathcal{G}(p_\parallel; x^D, y^D)$ and $\mathcal{G}_0(p_\parallel; x^D, y^D)$ as the total reduced propagator and the Feynman reduced propagator, respectively \cite{GTFABFEB,LHCBAFFFAB,Caval}. The reduced Feynman propagator $\mathcal{G}_0(p_\parallel; x^D, y^D)$ is explicitly given by
\begin{equation}
\label{reducedFeynmanProp}
    \begin{split}
        \mathcal{G}_0(p_\parallel;x^D,y^D)&=\int_{-\infty}^{+\infty}\frac{dp^D}{2\pi}\;\frac{e^{ip^D(x^D-y^D)}}{(p^D)^2+\sigma^2},\\
        &=\frac{e^{-\sigma|x^D-y^D|}}{2\sigma},
    \end{split}
\end{equation}
where $\sigma=\sqrt{m^2-p_\parallel^2}$.

By substituting equations (\ref{rprop1}) and (\ref{rprop2}) into (\ref{prop1}) and carrying out some algebraic manipulations, we obtain the following equation for the reduced propagator.
\begin{equation}
\label{prop2}
\mathcal{G}(p_\parallel;x^D,y^D)=\mathcal{G}_0(p_\parallel;x^D,y^D)-\frac{1}{M}\left[\partial_{a}\mathcal{G}(p_\parallel;x^D,a)\right]\left[\partial_a\mathcal{G}_0(p_\parallel;a,y^D)\right] \ .
\end{equation}

By setting $y^{D}=b$ in (\ref{prop2}), differentiating both sides with respect to $b$, and then evaluating at $b=a$ we obtain
\begin{equation}
\label{prop3}
\partial_a\mathcal{G}(p_\parallel;x^D,a) =\frac{ \partial_a\mathcal{G}_0(p_\parallel;x^D,a)} {1+\frac{1}{M}\partial_{a}\partial_{b}\left(\mathcal{G}_0(p_\parallel;a,b)\right)\Big|_{b=a}},
\end{equation}
where we used the fact that $\partial_a \mathcal{G}_0(p_\parallel;x^D,a)=\partial_b \mathcal{G}_0(p_\parallel;x^D,b)\Big|_{b=a}$, with the notation $\Big|_{b=a}$ indicating that the expression is evaluated at $b=a$.

With the insertion of (\ref{prop3}) in (\ref{prop2}), we are lead to the following equation 
\begin{equation}
\label{prop4}
\mathcal{G}(p_\parallel;x^D,y^D)=\mathcal{G}_0(p_\parallel;x^D,y^D) - \frac{ \left[\partial_a\mathcal{G}_0(p_\parallel;x^D,a)\right]\left[\partial_a\mathcal{G}_0(p_\parallel;a,y^D)\right]} {M+\partial_{a}\partial_{b}\left(\mathcal{G}_0(p_\parallel;a,b)\right)\Big|_{b=a}}.
\end{equation}

We now turn to discuss the result above. As seen in Eq. (\ref{prop4}), the total reduced propagator is expressed as the reduced Feynman propagator minus a correction term arising from the presence of the planar potential, which involves derivatives of the reduced Feynman propagator. It is important to note that the denominator of the second term on the right-hand side of Eq. (\ref{prop4}) is associated with the second-order derivative of the reduced Feynman propagator in Eq. (\ref{reducedFeynmanProp}) evaluated at the same point. A straightforward analysis indicates that this term diverges. To address this issue, we regularize and renormalize the expression. In this work, we use two regularization approaches. The first, which is employed here for the current discussion, involves the analytic continuation of integrals. The second approach uses the cutoff regularization method, which is discussed in detail in Appendix \ref{apen1}, where it is shown that both approaches yield the same results.

Using the the analytical extension of the following integral \cite{Kaku}
\begin{equation}\label{analyExten}
    \int_{0}^{\infty}dr\;\frac{r^\beta}{(r^2+c^2)^\alpha}=\frac{\Gamma(\frac{1+\beta}{2})\Gamma(\alpha-\frac{(1+\beta)}{2})}{2\Gamma(\alpha)(c^2)^{\alpha-(1+\beta)/2}},
\end{equation}
where $\Gamma(x)$ denotes the Euler Gamma function \cite{GBA}, we are able to regularize the second derivative of the reduced Feynman propagator given by the integral  (\ref{reducedFeynmanProp}), obtaining
\begin{equation}\label{secondDerivative}
    \begin{split}
       \partial_{a}\partial_{b}\mathcal{G}_0(p_\parallel;a,b)\Big|_{b=a} &=\frac{1}{\pi}\int_{0}^{\infty}dp\;\frac{p^\beta}
{(p^2+\sigma^2)^\alpha}\Bigg|_{\beta=2,\;\alpha=1},\\
        &=\frac{1}{\pi}\times\frac{\Gamma(\frac{1+\beta}{2})\Gamma(\alpha-\frac{(1+\beta)}{2})}{2\Gamma(\alpha)(\sigma^2)^{\alpha-(1+\beta)/2}}\Bigg|_{\beta=2,\;\alpha=1},\\
        &=-\frac{\sigma}{2}.
    \end{split}
\end{equation}
We note that above regularized expression is finite, and no further redefinition of parameters is needed in order to render finite the denominator in Eq.  (\ref{prop4}). In this way, by substituting (\ref{reducedFeynmanProp}) and (\ref{secondDerivative}) into Eq. (\ref{prop4}), performing the some calculations, and then inserting the resulting expression for the reduced total propagator into (\ref{rprop1}), we obtain the integral form of the propagator 
\begin{equation}
\label{propfinal}
 G(x,y)=\int \frac{d^{D}p_\parallel}{(2\pi)^D} \left[\frac{e^{-\sigma|x^D-y^D|}}{2\sigma}-\frac{sgn(x^D-a) sgn(y^D-a)}{2\left(2M-\sigma\right)}
e^{-\sigma\left(\mid x^D-a\mid +\mid a-y^D\mid\right)}  \right] e^{-ip_\parallel.(x_\parallel-y_\parallel)},
\end{equation}
where the sign function is defined as \( \text{sgn}(x) = 1 \) for \( x > 0 \), \( \text{sgn}(x) = -1 \) for \( x < 0 \), and \( \text{sgn}(0) = 0 \).

Theories involving planar \(\delta\)-like potentials possess the property of effectively generalizing boundary conditions, as noted in references \cite{PKKM,BorUM,BHR1992,GTFABFEB} for the Dirichlet boundary condition and in \cite{walisson} for the MIT boundary condition. Furthermore, there is the electromagnetic case, which considers an external potential that restores the propagator in the presence of a perfect conducting plate, as discussed in reference \cite{FABFEB}. For the model (\ref{model}), the situation is similar: the external potential effectively generalizes the Neumann boundary condition on the plane \(x^D = a\), where the Dirac delta in (\ref{model}) is localized, as we will demonstrate below. By applying the operator $n^\mu\partial_\mu=\partial_D$ in the propagator (\ref{propfinal}) and then evaluating the result at \(x^D = a\) we find
\begin{equation}
    \Big(n^\mu\partial_\mu G(x,y)\Big)\Big|_{x^D=a}=\int \frac{d^D p_\parallel}{(2\pi)^D}\;\Big(1-\frac{1}{1-2M/\sigma}\Big)\frac{sgn(y^D-a)e^{-\sigma|y^D-a|}}{2}e^{-ip_\parallel.(x_\parallel-y_\parallel)},
\end{equation}
then taking the limit \(M \to 0\), we obtain 
\begin{equation}\label{19}
    \begin{split}
        \lim\limits_{M\to 0}\Bigg[\Big(n^\mu\partial_{\mu} G(x,y)\Big)\Big|_{x^D=a}\Bigg]=0.\\
    \end{split}
\end{equation}
Thus, in the limit $M\to 0$ the Green's function of the model satisfies the Neumann boundary condition on the hyperplane $x^D=a$. Consequently, any classical field solutions must also satisfy this boundary condition.

\section{Interaction between a point charge  and the planar potential}
\label{interaction}

In this section, we compute the interaction energy between a stationary point-like source and the planar potential considered in Eq. (\ref{model}). We begin with the fact that the contribution to the system's energy arising from the external source $J(x)$ is given by \cite{GTFABFEB,LHCBAFFFAB,LHCAANEHFAB}
\begin{equation}
E=-\frac{1}{2T}\int d^{D+1}x\ d^{D+1}y\ J\left(x\right){{G}}\left(x,y\right)J\left(y\right),\label{energy}
\end{equation}
here, \(\int dx^0 = \int_{-T/2}^{T/2} dx^0\), where \(T\) denotes the time interval during which the external source is present, with the implicit limit \(T \to \infty\) taken at the end of the calculations.

We consider, without loss of generality, a point-like scalar charge placed at the position ${\bf B}=\left(0,\ldots,0,B^{D}\right)$, whose corresponding external source is given by
\begin{equation}
J\left(x\right)=\lambda\delta^{D}\left({\bf x}-{\bf B}\right),\label{source}
\end{equation}
where $\lambda$ is the scalar charge intensity. 

We notice that the first term on the right-hand side of Eq. (\ref{propfinal}) comes from the free Klein-Gordon
propagator (without the presence of the planar potential) and, therefore, it does not contribute to the interaction
energy between the scalar charge and the $\delta$-like potential. Indeed, this contribution does not depend on the distance between the 
charge and the planar potential, it is present even in the absence of the potential and provides the
charge self energy. So, only the second term of the propagator (\ref{propfinal}) contributes to the interaction energy.

Substituting Eqs. (\ref{source}) and (\ref{propfinal}) in (\ref{energy}), discarding self-interacting contributions, computing the integrals in the spacetime coordinates and carrying out some manipulations, we arrive at
\begin{equation}
\label{EMC1}
E_{int}=\frac{\lambda^2}{2(2\pi)^{D-1}} \int d^{D-1}\mathbf{p}_{\parallel} \frac{e^{-2R\sqrt{\mathbf{p}_{\parallel}^2+m^2}}}{4M-2\sqrt{\mathbf{p}_{\parallel}^2+m^2}} \ ,
\end{equation}
with $R=\mid B^{D}-a\mid$ standing for the distance between the scalar charge and the planar $\delta$-potential. The sub-index $int$ means 
that we have the interaction energy between the $\delta$-like potential and the charge.

The integral (\ref{EMC1}) can be simplified by using  hyperspherical coordinates \cite{GTFABFEB,FABGFH,Kaku}, as follows
\begin{equation}
\label{EMC2}
E_{int}=\frac{\lambda^2}{2(2\pi)^{D-1}} \ \Omega \int_{0}^{\infty} dr\frac{r^{D-2} e^{-2R\sqrt{r^2+m^2}}}{4M-2\sqrt{r^2+m^2}} \ ,
\end{equation}
where  the total solid angle $\Omega$ of the $(D- 1)$--sphere, is given by
\begin{equation}
\label{solid}
\Omega=\frac{2\pi^{\left(D-1\right)/2}}{\Gamma\left(\frac{D-1}{2}\right)}.
\end{equation}

Carrying out the change of variable $r\rightarrow y=2\sqrt{r^2+m^2}$ in (\ref{EMC2}) we arrive at 
\begin{equation}
\label{EMC3}
E_{int}\left(R,m,M,D\right)=\frac{\lambda^{2}}{4\left(4\pi\right)^{(D-1)/2}\Gamma\left((D-1)/2\right)}\int_{2m}^{\infty} dy\left[\frac{y^{2}}{4}-m^{2}\right]^{\frac{D-3}{2}}\frac{y \ e^{-Ry}}{4M-y}.
\end{equation}

The Eq. (\ref{EMC3}) is the general expression for the interaction energy between $\delta$-like external potential and a charge. In order to understand the  behavior
of this interaction, let us analyse separately the massless $m=0$ and massive $m\neq 0$ cases. 

\subsection{The massless case: $m=0$}

The first case of interest in which the integral in Eq. (\ref{EMC3}) can be solved analytically is the massless one. By setting $m=0$ in Eq. (\ref{EMC3}), we can write
\begin{equation}
\label{EMC5}
E_{int}\left(R,m=0,M,D\right)=\frac{\left(-1\right)^{D}\lambda^{2}}{\left(16\pi\right)^{(D-1)/2}\Gamma\left((D-1)/2\right)}\frac{d^{D-2}}{dR^{D-2}}\int_{0}^{\infty} dy \frac{e^{-Ry}}{4M-y}.
\end{equation}
Now we analyze separately the cases $M=0$, $M<0$ and $M>0$.

\subsubsection{The masless Neumann case: $M=0$}

As discussed previously, when $M=0$, the model corresponds physically to a scalar field subjected to the Neumann boundary conditions at the plane $x^D=a$. In this case, by setting $M=0$ in (\ref{EMC5}) we readily obtain
\begin{equation}  
E_{int}\left(R,m=0,M=0,D\right)=\frac{(-1)^D\lambda^2}{4^{D-1}\pi^{\frac{D-1}{2}}\Gamma(\frac{D-1}{2})}
\frac{\partial^{D-3}}{\partial R^{D-3}}R^{-1}\ .
\label{Mm00}
\end{equation}

This expression is the negative of the one found in reference \cite{GTFABFEB} for the generalized Dirichlet boundary condition. It indicates that the presence of the Neumann mirror, in the presence of the scalar charge, is equivalent to an image scalar charge with the same signal of the real one.

For $D=3$ spatial dimensions, Eq. (\ref{Mm00}) reads
\begin{equation}  
E_{int}\left(R,m=0,M=0,D=3\right)=-\frac{\lambda^2}{16\pi R},
\label{Mm003}
\end{equation}
which corresponds to the standard attractive Coulomb-like interaction between a point-like source and its image. It is worth noting that, for the Klein–Gordon field, scalar charges with the same sign attract each other.

\subsubsection{The masless case with negative coupling: $M<0$}

Replacing $M\rightarrow -\mid M\mid$ and performing the change of variable $y\rightarrow z=\frac{\left(y+4 \mid M\mid\right)}{4\mid M\mid}$ in Eq. (\ref{EMC5}) we have
\begin{equation}
\label{EMC7}
E_{int}\left(R,m=0,M<0,D\right)=\frac{\left(-1\right)^{D}\lambda^{2}}{\left(16\pi\right)^{(D-1)/2}\Gamma\left((D-1)/2\right)}\frac{d^{D-2}}{dR^{D-2}}\left\{e^{-4 |M| R}
 Ei\left(-4|M| R\right)\right\},
\end{equation}
where $Ei\left(x\right)$ denotes the exponential integral function \cite{GBA},
\begin{equation}
\label{Eix}
Ei\left(x\right)=-\int_{-x}^{\infty}\frac{e^{-t}}{t}dt=\int_{-\infty}^{x}\frac{e^{t}}{t}dt  ,
\end{equation}
which is valid for $x<0$. For $x>0$, the integral is interpreted in the sense of the Cauchy principal value.

For concreteness, we take the case $D=3$ in (\ref{EMC7}), which yields the following expression for the interaction energy
\begin{equation}\label{m0D3}
    E_{int}(R,m=0,M<0,D=3)=\frac{-|M|\lambda^2}{4\pi}\Bigg[\frac{1}{4|M|R}+e^{4|M|R}Ei(-4|M|R)\Bigg]
\end{equation}

Note that the interaction energy (\ref{m0D3}) has two contributions. The first term is an attractive Coulomb-like interaction energy,
independent of parameter $M$. The second term, that depends on the parameter $M$, gives a repulsive contribution. The first term dominates over the second one for any value of $M<0$ and for any distance $R$.

\begin{figure}[H]
    \centering
    \includegraphics[scale=0.7]{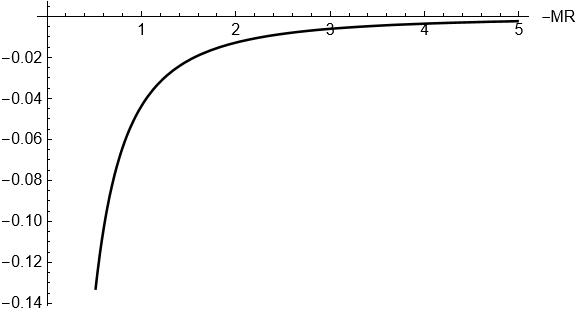}
    \caption{Interaction energy (\ref{m0D3}) multiplied by $\frac{4\pi}{\lambda^2|M|}$ as a function of $|M| R$ in the $D=3$ scenario.}
    \label{energia m=0 M<0 real}
\end{figure}

In FIG. \ref{energia m=0 M<0 real} we present a plot of the rescaled interaction energy, $E_{int}\left(R,m=0,M<0,D=3\right)\times[4\pi/(|M|\lambda^{2})]$, obtained from (\ref{m0D3}), as a function of $|M|R$. The plot reveals an attractive interaction between the point-like source and the planar potential, with the interaction energy decreasing as $R$ increases, as expected.

\subsubsection{The masless case with positive coupling: $M > 0$}

As discussed just below Eq. (\ref{model}), a small imaginary part is implicitly included in the field mass, $m^{2}\to m^{2}-i\epsilon$. Taking this into account in the denominator of the integral in Eq. (\ref{EMC5}) is equivalent to the prescription $y\rightarrow y-i\epsilon$, which ensures the convergence of the integral for $M>0$. Accordingly, for \( M > 0 \), we explicitly adopt the prescription $y\rightarrow y-i\epsilon$ in the integral on the right-hand side of Eq. (\ref{EMC5}). At the end of the calculation, we must take the limit $\epsilon\rightarrow 0$. Thus
\begin{eqnarray}
\label{zxc2}
\int_{0}^{\infty} dy \frac{e^{-Ry}}{4M-y}
\to \lim_{\epsilon\to 0^+}\int_{0}^{\infty} dy \frac{e^{-Ry}}{4M-y+i\epsilon}\ .
\end{eqnarray}

By using the identity
\begin{equation}
\lim_{\epsilon\to 0^+} \frac{1}{x-i\epsilon}={\cal P}\left(\frac{1}{x}\right)+i\pi\delta(x)
\end{equation}
we obtain
\begin{equation}
\label{intM0}
\lim_{\epsilon\to 0^+}\int_{0}^{\infty} dy \frac{e^{-Ry}}{4M-y+i\epsilon}
=e^{-4MR}\left[Ei\left(4MR\right)-i\pi\right], 
\end{equation}
where  $Ei\left(x\right)$ is given by the Cauchy principle value of integral (\ref{Eix}). 
Using (\ref{intM0}) in (\ref{EMC5}) we have
\begin{equation}
\label{EMC8}
E_{int}\left(R,m=0,M>0,D\right)=\frac{\left(-1\right)^{D}\lambda^{2}}{\left(16\pi\right)^{(D-1)/2}\Gamma\left((D-1)/2\right)}\frac{d^{D-2}}{dR^{D-2}}\Bigl\{e^{-4MR}\left[Ei\left(4MR\right)-i\pi\right]\Bigr\},
\end{equation}

It is interesting to notice that for $M>0$ the interaction energy (\ref{EMC8}) exhibits a negative imaginary part for every spatial dimension $D\geq 1$, which is given by
\begin{equation}\label{imaginaryOfEMC8}
   {\rm  Im}\Big\{E_{int}(R,m=0,M>0,D)\Big\}=-\frac{\lambda^2\;M^{D-2}}{4 \pi^{\frac{D-3}{2}}\Gamma((D-1)/2)}\;e^{-4M R}.
\end{equation}

The imaginary part above can be interpreted in terms of the vacuum-to-vacuum transition amplitude of the field in the presence of both the external potential and the external source,
\begin{equation}
\label{zxc1}
    \langle 0|0\rangle _J = e^{-i T E}\ .
\end{equation}
Therefore, if we have
$E ={\rm Re}(E)+ i\;{\rm Im}(E)$ with ${\rm Im}(E)<0$,
the vacuum-to-vacuum transition (\ref{zxc1}) decreases exponentially with time, as follows
\begin{equation}\label{vacuumexp}
    \langle 0|0\rangle _J= e^{-i T {\rm Re}(E)}\; e^{T {\rm  Im}(E)}.
\end{equation}

In this way, the physical consequence of a negative imaginary part of the energy is the instability of the vacuum state. 
For the model (\ref{model}), this instability arises from the interaction between the point-like source and the planar potential, and it depends on the coupling parameter $ M $, the charge $ \lambda $ of the source, and the distance $ R $ between the source and the planar 
potential, as shown in (\ref{imaginaryOfEMC8}).

As the vacuum-to-vacuum transition amplitude decreases, other particle states become occupied. We interpret $( 2|\text{Im}(E)|=2 |{\rm  Im}(E_{int})| )$ as the particle-antiparticle creation rate, analogous to the Schwinger pair creation effect in Quantum Electrodynamics, which arises due to the presence of an external electric field \cite{Schwinger1950}. In the present model, the point-like source plays the role of the constant external electric field, in the presence of the potential.

Denoting the particle-antiparticle creation rate (or equivalently the vacuum decay rate) as $\Gamma_v$ and using
(\ref{imaginaryOfEMC8}), we have
\begin{equation}
\Gamma_v=\frac{\lambda^2\;M^{D-2}}{2\pi^{\frac{D-3}{2}}\Gamma((D-1)/2)}\;e^{-4M R},
\label{decay}
\end{equation}
which exhibits an exponential decreasing behavior as function of $MR$. In this way, the pair production rate is suppressed
for sufficiently large distance $R$ between the planar potential and the point-like source.

The real part of expression (\ref{EMC8}) measures the interaction energy between the planar potential and the point-like source,
\begin{equation}
\label{EMRe}
{\rm Re}\Big\{E_{int}(R,m=0,M>0,D)\Big\}=\frac{\left(-1\right)^{D}\lambda^{2}}{\left(16\pi\right)^{(D-1)/2}\Gamma\left((D-1)/2\right)}\frac{d^{D-2}}{dR^{D-2}}\Bigl\{e^{-4MR}Ei\left(4MR\right)\Bigr\}.
\end{equation}

For $D=3$, expression (\ref{EMRe}) reads
\begin{equation}
{\rm Re}\Big\{E_{int}(R,m=0,M>0,D=3)\Big\}=\frac{\lambda^2M}{4\pi}\left[-\frac{1}{4MR}+e^{-4MR}Ei(4MR)\right],
\label{EMReD3}
\end{equation}
which has the same functional form as the corresponding one for the case $M<0$, given by (\ref{m0D3}).

\begin{figure}[H]
   \centering
 \includegraphics[scale=0.6]{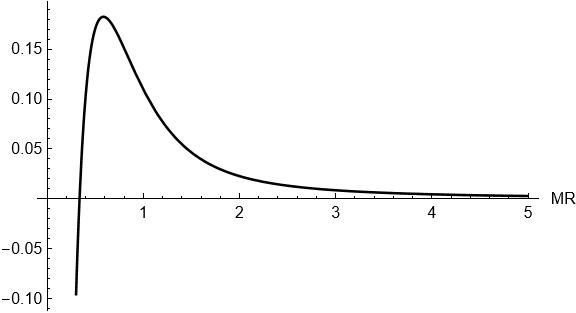}
   \caption{The interaction energy (\ref{EMReD3}),
 multiplied by \(\frac{4\pi }{\lambda^2 M}\) for the massless case $m=0$ and positive coupling constant $M>0$, as a function of \(MR\) in the \(D = 3\) scenario.}
    \label{energia m=0 M>0 real MR}
\end{figure}

The behavior of the interaction energy (\ref{EMReD3}) is shown in FIG. \ref{energia m=0 M>0 real MR} as function of $MR$. We highlight a peak at $MR\sim0.58$, corresponding to a point of unstable equilibrium, where the force vanishes.  From this point onward, the force between the charge and the potential becomes repulsive. In the region between the potential and the peak, the interaction is attractive and diverges as the charge approaches the potential.

\subsection{The massive case, $m\neq 0$}

For convenience, we separate the analysis of the massive case $m\not=0$ for the situations  $M=0$, $M<m/2$ and $M> m/2$.

\subsubsection{Massive Neumann case: $m\not=0$, $M=0$}
 In this situation the expression (\ref{EMC3}) reads
\begin{equation}
\label{EMC9}
E_{int}\left(R,m,M=0,D\right)=-\frac{\lambda^{2}}{4\left(4\pi\right)^{(D-1)/2}\Gamma\left((D-1)/2\right)}\int_{2m}^{\infty} dy\left[\frac{y^{2}}{4}-m^{2}\right]^{(D-3)/2}e^{-Ry},
\end{equation}
which, for $D=3$,
\begin{equation}
E_{int}\left(R,m,M=0,D=3\right)=-\frac{\lambda^{2}}{16\pi}\frac{ e^{-2mR}}{R}.
\label{mM0D3}
\end{equation}

The result in (\ref{mM0D3}) describes a Yukawa-like interaction between the charge and its image, arising from the two-dimensional $\delta$-like planar potential. It is important to highlight, however, that this interaction is attractive—contrary to the case of a Dirichlet-like planar potential, where the interaction is repulsive \cite{GTFABFEB}. This contrast suggests that, in the case of a Neumann-like planar potential, the image charge has the same sign as the original charge, unlike the Dirichlet-like scenario, where the image charge has the opposite sign \cite{GTFABFEB}.

For a general value of $D$, we get for (\ref{EMC9}) 
\begin{eqnarray}
\label{EMC10}
E_{int}\left(R,m,M=0,D\right)=-\frac{\lambda^{2}m^{D-2}}{2\left(2\pi\right)^{D/2}}\left(2mR\right)^{1-(D/2)}K_{(D/2)-1}\left(2mR\right),
\end{eqnarray} 
where $K_{n}\left(x\right)$ stands for the $K$-Bessel function \cite{GBA}. The result in Eq. (\ref{EMC10}) is the same as that obtained for two scalar charges separated by a distance $2R$ in a $(D+1)$-dimensional spacetime \cite{FABGFH}. Therefore, we can interpret the Neumann limit effectively as the interaction between the real charge and its image. By setting $D=3$ in Eq. (\ref{EMC10}), we obtain Eq. (\ref{mM0D3}).

\subsubsection{Massive case with $M<m/2$}

\begin{figure}[H]
\centering \includegraphics[scale=0.6]{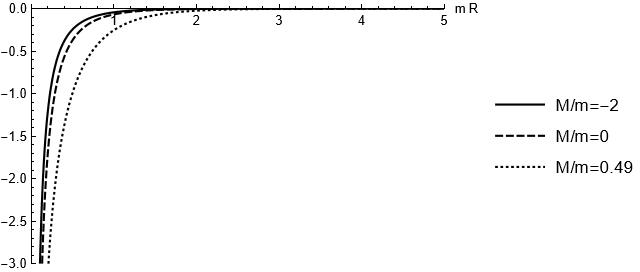}
\caption{Plot of  interaction energy (\ref{EMC31}) multiplied by $\frac{4\pi}{\lambda^{2}m}$ as a function of $mR$ when 
$M/m=-2$ (solid line), $M/m=0$ (dashed line) and $M/m=0.49$ (dotted line).}
\label{grafico2MR}
\end{figure}

To gain better insight, we start by considering a \((3+1)\)-dimensional spacetime, corresponding to \(D = 3\). In this scenario, expression (\ref{EMC3}) can be easily integrated
\begin{equation}
\label{EMC31}
E_{int}\left(R,m,M<m/2,D=3\right)=\frac{\lambda^{2} m}{4\pi}
\left[-\frac{e^{-2mR}}{4mR}+\frac{M}{m} \ e^{-4M R}Ei\left(4MR-2m R\right)\right],
\end{equation}
where in the first term on the right hand side we have the attractive Yukawa-like potential (\ref{mM0D3}) and a correction due to the non-zero coupling constant $M^{-1}$. Such a correction is of repulsive character for $M<0$, attractive for $M>0$ and  falls down when the distance $R$ increases as well as when the parameter $|M|$ decreases. In any case, the net result is of an attractive character for all $R$.
 To illustrate this discussion, in FIG. \ref{grafico2MR} we display the behavior of the  interaction energy (\ref{EMC31}) multiplied by $\frac{4\pi}{\lambda^{2}m}$, as a function of $mR$, for some representative values of $M<m/2$, including the Yukawa-like interaction $M=0$.
We observe an attractive  interaction energy for all distances $R$. 

\subsubsection{Massive case with $M>m/2$}

Restricted to the case $D=3$, and proceeding as in the massless case with $M>0$ (see the discussion preceding Eq. (\ref{zxc2})), we explicitly use the prescription \( y \rightarrow y - i\epsilon \) in the denominator of the integral (\ref{EMC3}). In this way, we readily obtain
\begin{eqnarray}
\label{EMC33}
E_{int}\left(R,m,M>\frac{m}{2},D=3\right)=
\frac{\lambda^{2}m}{4\pi}\left[-\frac{e^{-2mR}}{4m R}+\frac{M}{m}e^{-4M R}E_{i}(4M R-2mR)-i\pi\frac{M}{m}e^{-4M R} \right],
\end{eqnarray}
where we have a negative imaginary part for the energy, which remains unchanged as in the massless case $m=0$, $M>0$ and given by 
by Eq. (\ref{imaginaryOfEMC8}) for $D=3$. Consequently we have vacuum instability with decay rate given by Eq. (\ref{decay})
with $D=3$. On the other hand, the force interaction between the point source and the planar potential is related to the real part of expression (\ref{EMC33}), 
\begin{equation}
{\rm Re}\left\{E_{int}\left(R,m,M>\frac{m}{2},D=3\right)\right\}=
\frac{\lambda^{2}m}{4\pi}\left[-\frac{e^{-2mR}}{4 m R}+\frac{M}{m}\ e^{-4MR}Ei\left(4MR-2mR\right)\Bigr) \right],
\label{EMC33b}
\end{equation}
which consists of an attractive Yukawa-like interaction plus a second term that incorporates the effect of the planar potential. 
The second term can be of an attractive or repulsive character depending on the dimensionless distance $mR$. The net result for
the  interaction energy behavior as function of distance, $R$, is similar to the massless case $m=0$ with $M>0$, attrative up to a peak
and repulsive from it.

\begin{figure}[H]
\centering 
\includegraphics[scale=0.7]{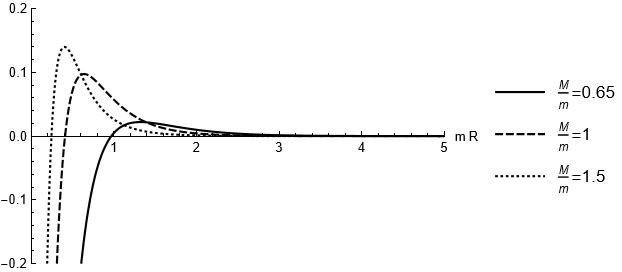} 
\caption{Plot of interaction energy (\ref{EMC33b}) multiplied by $\frac{4\pi}{\lambda^2 m}$ as a function of $mR$ for $M/m=0.65$ 
(solid line), $M/m=1$ (dashed line) and $M/m=1.5$ (dotted line).}
\label{grafico4MR}
\end{figure}

\begin{figure}[H]
\centering 
\includegraphics[scale=0.7]{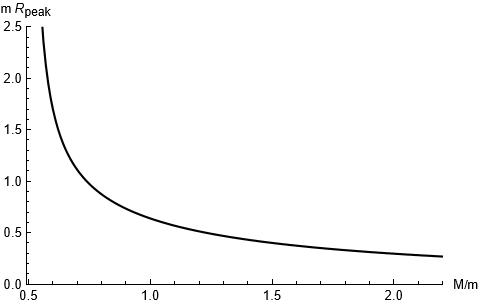} 
\caption{Position of the peaks multiplied by $m$, as a function of $M/m$.}
\label{picos}
\end{figure}

To illustrate this behavior, we plot in FIG.\ref{grafico4MR} the interaction energy given by Eq. (\ref{EMC33b}), multiplied by $\frac{4\pi}{\lambda^2 m}$, as a function of $mR$ for several values of the ratio $M/m>1/2$. We observe that the position of the peak increases as $M/m$ decreases, and vice versa. Conversely, the maximum value of the interaction energy at the peak decreases as $M/m$ decreases.In the limit $M/m\to 1/2$, it appears that the peak vanishes, as illustrated in FIG. \ref{picos}, where we plot the peak position multiplied by $m$, $mR_{peak}$, as a function of $M/m$.
When $M=\infty$, corresponding to the Neumann limit, the peak shifts to infinity, as expected.

The graphic FIG (\ref{picos}) exhibits an asymptote at $M/m=1/2$, as expected, because below this ratio the interaction energy is always monotonic, and there is no vacuum decay.

To the best of the authors' knowledge, the presence of a local maximum or minimum in the interaction energy is a feature unique to this model, with no counterpart in previously studied models, even in theories involving higher-order derivatives and external potentials \cite{RodriguesNogueiraBarone2024}, where the force may exhibit a local minimum but retains the same sign, and the energy remains monotonic.

A standard situation in which a similar mechanism of vacuum decay occurs can be found in QED \cite{GreinerQED}. However, in that case, there is no external potential, but rather an external field — the electromagnetic field - whose magnitude exceeds a critical value. This field couples to the quantum fermionic field, whose vacuum becomes unstable under such conditions.

\subsection{The singular case, $M=m/2\not=0$ and the general result}

For the case where $M=m/2\not=0$, the integral in Eq. (\ref{EMC3}) is divergent. This behavior is expected, as the integral in Eq. (\ref{propfinal}) is also divergent. As a result, the Green's function of the system is ill-defined, and the model becomes singular in this situation. This indicates that the model is not physically well-defined for this particular case.

In fact, when $M=m/2$, a logarithmic divergence arises from the $Ei(x)$ function in all expressions where $M\not=0$. It can be seen from the expansions
\begin{equation}
Ei(\pm x)\approx \gamma+\ln(x)\pm x+{\cal O}(x^2),
\end{equation}
with $\gamma$ standing for the Euler constant.

This divergence depends on the parameter $R$ and cannot be removed through renormalization or with the Feynman prescription for the mass of the field, $m\to m-i\epsilon$. Conversely, when $M=m=0$, all terms involving the $Ei(x)$ function vanish, and the interaction energy becomes well-defined.

All the results obtained for the interaction energy between the planar potential and the point-like scalar charge in $D=3$ can be summarized in the following single expression,
\begin{eqnarray}
\label{energiaGeral}
E_{int}\left(R,m,M,D=3\right)=
\frac{\lambda^{2}}{4\pi}\left[-\frac{e^{-2mR}}{4 R}+Me^{-4M R}E_{i}(4M R-2mR)-i\pi Me^{-4M R}\Theta\bigg(M-\frac{m}{2}\bigg)\right]\ ,\cr\cr
\ \mbox{if } M-(m/2)\not=0\ \mbox{for }\ M\not=0.
\end{eqnarray}
where $\Theta(x)$ stands for the Heaviside function, defined by
\begin{equation}
 \Theta(x) = \begin{cases}
1, & x > 0 \\
\frac{1}{2}, & x = 0 \\
0, & x < 0
\end{cases}\ .
\end{equation}


\section{Conclusions and final remarks}
\label{conclusions}

The massive scalar field theory in $D+1$ dimensions, in the presence of an external planar potential that couples quadratically to the normal derivative of the field — as described by the Lagrangian density in Eq. (\ref{model}) — has been investigated. We have determined exactly how the presence of the $\delta$-like planar potential modifies the Feynman propagator. In the appropriate limit of the coupling parameter between the field and the planar potential, namely $M\to0$, the propagator satisfies the Neumann boundary condition on the plane. Conversely, in the opposite limit, $M\to\pm\infty$, the coupling between the field and the potential vanishes, and the propagator reduces to that of the standard Klein-Gordon Lagrangian.

We have calculated the interaction energy between a stationary point-like scalar source and the planar potential, as given in general by Eq. (\ref{EMC33b}), focusing on the case $D=3$. We found that when the coupling constant $M$ and the Klein-Gordon mass $m$ satisfy the condition 
$M<m/2$, the interaction energy is always attractive, regardless of the distance $R$ between the source and the potential.

On the other hand, when $M>m/2$, the interaction energy becomes non-monotonic and exhibits mixed behavior: it is attractive up to a certain distance - corresponding to a peak where the energy reaches a maximum - and repulsive beyond that point. This peak represents an unstable equilibrium position.

Moreover, still under the condition $M>m/2$, we have shown that the vacuum state of the system becomes unstable in the presence of the stationary point-like source. This instability triggers a pair production process analogous to the Schwinger effect in quantum electrodynamics, which occurs in the presence of a constant external electric field with a magnitude exceeding a critical value. In the model (\ref{model}), the analogous role of the electric field is played by the stationary point-like source. We found that the vacuum decay rate is independent of the field mass $m$ and is given by Eq. (\ref{decay}) in the $D=3$ case.

We hope that the results presented here may contribute to the understanding of the role that external potentials coupled to fields can play in the description of material boundaries. It is worth noting that many condensed matter systems can be effectively modeled using scalar fields, especially those related to phonons or magnons. Scalar field models are also commonly used as toy models for the electromagnetic field.

The model studied in this work could be considered in several other contexts, such as the presence of two parallel planar potentials - potentially leading to a Casimir effect and a modified decay rate - or scenarios involving a moving potential, the propagation of scalar field waves in the presence of the planar potential, and so on. We leave these as open questions for interested readers.

We have calculated the interaction energy between a stationary point-like scalar source and the planar potential, given in general by Eq.
(\ref{EMC33b}), restricted to $D=3$ scenarios. We have found that when the coupling constant $M$ and the Klein-Gordon mass $m$ satisfy the relation $M<m/2$, the interaction energy between the source and the planar potential is attractive for all values of the distance $R$ between the source and the potential. 

On the other hand, when $M>m/2$, the interaction energy is no longer monotonic and instead exhibits a mixed behavior: it is attractive up to a certain distance—where it reaches a maximum—and becomes repulsive beyond that point. This peak corresponds to an unstable equilibrium position.

Furthermore, still under the condition $M>m/2$, we have shown that the vacuum state of the system becomes unstable in the presence of the stationary point-like source. This instability gives rise to a pair production process, analogous to the Schwinger effect in Quantum Electrodynamics, where an external constant electric field exceeding a critical value leads to spontaneous pair creation. In our model, the stationary point source plays a role analogous to that of the electric field in QED. We have found that the vacuum decay rate is independent of the field mass $m$ and is given by expression (\ref{energiaGeral}) in the case $D=3$.

\begin{acknowledgments}
For financial support, F.A. Barone acknowledges CNPq under grant 313426/2021-0. J.P. Ferreira acknowledges support from CNPq, and L.H.C. Borges acknowledges support from CNPq and FAPEMIG under grant APQ-06536-24.
\end{acknowledgments}

\appendix
\section{Cut-off regularization procedure }\label{apen1}
Here we will show the cut-off regularization procedure approach for the result given by (\ref{propfinal}). In this way we have
for the  second derivative of the reduced Feynman propagator,
\begin{eqnarray}
\label{cutreg1}
 \lim\limits_{b\to a}\partial_b \partial_a \mathcal{G}_{0}(P_\parallel;a,b)&=&
 \frac{1}{\pi}\int_{0}^{\Lambda}dP\frac{P^2}{\sigma^2+P^2}\nonumber\\
&=&\frac{\sigma}{\pi}\Big( \frac{\Lambda}{\sigma}-\frac{\pi}{2}\Big)
\end{eqnarray}
where  $\Lambda\to\infty$. Note, that different form the analytical regularization method, the above regularized expression diverges. To eliminate
such divergence, we redefine the parameters of the model. As will be showed below, for such purpouse, it is sufficient to redefine the inverse coupling parameter $M$ in (\ref{model}). In terms of renormalized $M_R$, we write $M=Z_M M_R$. Consequently, we have for the reduced propagator
(\ref{prop4}),
\begin{equation}\label{renormg}
    \begin{split}
        \mathcal{G}(P_\parallel;x^D,y^D)= \mathcal{G}_{0}(P_\parallel;x^D,y^D) - 
\frac{\partial_a \mathcal{G}_{0}(P_\parallel;x^D,a)\partial_a \mathcal{G}_{0}(P_\parallel;a,y^D)}{Z_M M_R + \lim\limits_{b\to a}\partial_b \partial_a \mathcal{G}_{0}(P_\parallel;a,b)}.
    \end{split}
\end{equation}
where now, in the denominator we have the regularized expression for the second derivative of the reduced Feynman propagator. Using
(\ref{cutreg1}) in  above expression, we have 
\begin{equation}
    \begin{split}
        Z_M M_R + \lim\limits_{b\to a}\partial_b \partial_a \mathcal{G}_{R,0}(P_\parallel;a,b)&= Z_M M_R +  \frac{\Lambda}{\pi}-\frac{\sigma}{2},
    \end{split}
\end{equation}
and choosing $Z_M=1-\frac{\Lambda}{\pi M_R}$ we eliminate the divergence in above expression and find the same result,
Eq. (\ref{propfinal}) as obtained by the analytical regularization, where the parameter $M$ should be understood as the renormalized one.

\end{document}